\documentclass[fleqn,twoside]{article}

\usepackage[headings]{espcrc2}

\def\small{}

\newskip\humongous \humongous=0pt plus 1000pt minus 100pt
\def\caja{\mathsurround=0pt}
\def\eqalign#1{\,\vcenter{\openup1\jot \caja
       \ialign{\strut \hfil$\displaystyle{##}$&$\displaystyle{{}##}$\hfil\crcr#1\crcr}}\,} 
\newif\ifdtup

\newcounter{eqnumber}[section]
\renewcommand{\theeqnumber}{\thesection.\arabic{eqnumber}}
\def\equn{\refstepcounter{eqnumber}
\eqno({\rm \theeqnumber})
}

\def\npb#1#2#3{{\rm Nucl. Phys. B}{\bf \ #1}, #3 (#2)}

\def\cqg#1#2#3{{\rm Class. and Quant.\ Grav.} {\bf  #1}, #3 (#2)}

\def\hepth#1{[hep-th/#1]}
\def\hepph#1{[hep-ph/#1]}

\newbox\charbox
\newbox\slabox
\def\s#1{{      
        \setbox\charbox=\hbox{$#1$}
        \setbox\slabox=\hbox{$/$}
        \dimen\charbox=\ht\slabox
        \advance\dimen\charbox by -\dp\slabox
        \advance\dimen\charbox by -\ht\charbox
        \advance\dimen\charbox by \dp\charbox
        \divide\dimen\charbox by 2
        \raise-\dimen\charbox\hbox to \wd\charbox{\hss/\hss}
        \llap{$#1$}
}}

\def\spa#1.#2{\left\langle#1\,#2\right\rangle}
\def\spb#1.#2{\left[#1\,#2\right]}
\def\lor#1.#2{\left(#1\,#2\right)}

\catcode`@=11  

\def\NP{{\rm NP}}

\def\eps{\epsilon}

\def\I{{\cal I}}

\def\twoloop{{2 \mbox{-} \rm loop}}

\def\spa#1.#2{\left\langle#1\,#2\right\rangle}
\def\spb#1.#2{\left[#1\,#2\right]}
\def\lor#1.#2{\left(#1\,#2\right)}

\def\sand#1.#2.#3{%
  \left\langle\smash{#1}{\vphantom1}\right|{#2}%
  \left|\smash{#3}{\vphantom1}\right\rangle}
\def\sandp#1.#2.#3{%
  \left\langle\smash{#1}{\vphantom1}^{-}\right|{#2}%
  \left|\smash{#3}{\vphantom1}^{+}\right\rangle}
\def\sandpp#1.#2.#3{%
  \left\langle\smash{#1}{\vphantom1}^{+}\right|{#2}%
  \left|\smash{#3}{\vphantom1}^{+}\right\rangle}
\def\sandmm#1.#2.#3{%
  \left\langle\smash{#1}{\vphantom1}^{-}\right|{#2}%
  \left|\smash{#3}{\vphantom1}^{-}\right\rangle}
\def\sandpm#1.#2.#3{%
  \left\langle\smash{#1}{\vphantom1}^{+}\right|{#2}%
  \left|\smash{#3}{\vphantom1}^{-}\right\rangle}
\def\sandmp#1.#2.#3{%
  \left\langle\smash{#1}{\vphantom1}^{-}\right|{#2}%
  \left|\smash{#3}{\vphantom1}^{+}\right\rangle}

\def\Aloop{A^{\rm 1-loop}}

\def\Mloop{M^{\rm 1-loop}}
\def\Mtree{M^{\rm tree}}

\def\Atree{A^{\rm tree}}
\def\Aloop{A^{\rm 1-loop}}

\def\tree{{\rm tree}}

\def\NeqEight{{\cal N} = 8}
\def\NeqFour{{\cal N} = 4}

\title{Perturbative Gravity and Twistor Space}

\author{
N.~E.~J.~Bjerrum-Bohr\address[UWS]{Department of Physics, 
University of Wales Swansea}\ \address[comment]{Presented by N. E. J. 
Bjerrum-Bohr at Loop and Legs 2006
},
David~C.~Dunbar\addressmark[UWS]
and
Harald~Ita\addressmark[UWS] SWAT-06/464
}

\begin{document}

\maketitle

\section{Introduction}

Tree amplitudes in gravity theories can be computed from those
of gauge theory using the KLT relations~\cite{KLT}. Inspired by this 
observation, investigations~\cite{BDDPR,Bern:1998sv,BeBbDu,BBDuIt} have been carried out in the hope of 
discovering a deeper connection at the perturbative 
level between $\NeqEight$ supergravity~\cite{ExtendedSugra} 
and $\NeqFour$ super Yang-Mills. 
A link between these theories, beyond structural similarities 
like the non-abelian gauge symmetry and maximal supersymmetry, is not 
obvious and could have surprising implications on the UV behaviour 
of $\NeqEight$ supergravity.
After all, $\NeqFour$ super Yang-Mills can be proven to be a finite 
theory, to all orders in perturbation theory~\cite{Mandelstam}. 
Explicit computations seem to be the way forward in order to derive
concrete statements about $\NeqEight$ supergravity.  

The ``weak-weak'' duality, between $\NeqFour$ super Yang-Mills 
and a topological string theory propagating in twistor 
space~\cite{Witten:2003nn} implies the existence of a single 
perturbative $S$-matrix for these two theories.
Within the $S$-matrix of gauge theory the duality lead to the 
discovery of surprising structures as the MHV-vertex 
construction~\cite{CSW,CSWLoop} and the BCFW recursion 
relations~\cite{Britto:2004ap}. 
Although the duality applies more readily to $\NeqFour$ 
SYM~\cite{NeqFour} the ideas motived by the twistor space 
duality have been applied successfully to a much wider 
range of theories and in particular to gravity. 
Among other applications are theories both with less 
supersymmetry~\cite{NeqOne}, massive particles~\cite{CSW:matter,CSW:massive} 
and computations of QCD one-loop amplitudes~\cite{OneLoopQCD}.

In this talk we discuss how the ideas inspired by the 
``weak-weak'' duality can be applied to calculate 
one-loop $\NeqEight$ supergravity amplitudes.

We interprete the result to give evidence for the 
the "no-triangle" hypothesis, which states 
that $\NeqEight$ supergravity contains neither 
triangle nor bubble scalar integral functions.
The absence of these integral functions implies 
the de-facto power counting to be similar to that of $\NeqFour$ SYM and 
thus stronger than needed for finiteness at one-loop. 
Such simplifications require, if they extend beyond one-loop, 
a rethinking of the ultra-violet structure of 
maximal supergravity.

\section{Tree Amplitudes in Gravity Theories} 
Tree amplitudes in gravity theories are linked to those of 
gauge theory~\cite{KLT} via the heuristic relation 
in string theory,
\vspace{-0.08cm}
$$\vspace{-0.15cm}
\eqalign{\begin{pmatrix}{\rm closed\cr \rm string}\end{pmatrix} & \sim  \begin{pmatrix}{\rm open\cr \rm string}\end{pmatrix}_{\rm Left}\hspace{-0.15cm}\times \begin{pmatrix}{\rm open\cr \rm string}\end{pmatrix}_{\rm Right}
\cr} 
\hspace{-1.45cm}\equn\label{StringRelation}
$$\vspace{-.12cm}\\
The concrete realisation of the relationship
up to six points at $\alpha'=0$ is,
\vspace{-0.08cm}
$$\hspace{-0.2cm}
\eqalign{
M_{[1,2,3]}^{\rm tree} &\!=
-i\,A_{[1,2,3]}^{\rm tree}\!\times\! A_{[1,2,3]}^{\rm tree}\,,
\cr
M_{[1,2,3,4]}^{\rm tree}&\!=
-i\,s_{12}\,A_{[1,2,3,4]}^{\rm tree}\!\times\! A_{[1,2,4,3]}^{\rm tree}\,, \label{KLTFour} \cr
M_{[1,2,3,4,5]}^{\rm tree}&\!=
\ i\,s_{12}s_{34}\, A_{[1,2,3,4,5]}^{\rm tree}\!\times\! A_{[2,1,4,3,5]}^{\rm tree} \cr 
& +i s_{13}s_{24}\, A_{[1,3,2,4,5]}^{\rm tree}\!\times\! A_{[3,1,4,2,5]}^{\rm tree}\,,
\label{KLTFive} \cr
M_{[1,2,3,4,5,6]}^{\rm tree}&\!=
\!\! -i  s_{12}s_{45}\, A_{[1,2,3,4,5,6]}^{\rm tree}\!\times\!\big[s_{35}A_{[2,1,5,3,4,6]}^{\rm tree}
\cr   & \hspace{-0.1cm}
+(s_{34}+s_{35})  \ A_{[2,1,5,4,3,6]}^{\rm tree}\big ] + {\cal P}(2,3,4)\,,
\cr}
\hspace{-1.5cm}\equn\label{KLTSix}
\vspace{-0.2cm}
$$
where
$s_{ij} = (k_i + k_j)^2$, ${\cal P}(2,3,4)$ represents 
the sum over permutations of legs $2,3,4$ and the 
$A^\tree_n$ are tree-level colour-ordered gauge theory partial
amplitudes. These relations are the Kawai, Lewellen 
and Tye (KLT) relations~\cite{KLT}. Even in low energy
effective field theories for gravity~\cite{Donoghue:1994dn,EffKLT}
the KLT-relations can be seen to be valid.
 
For Yang-Mills amplitudes, the twistor space duality has 
motivated the development of a MHV-vertex reformulation 
for tree amplitudes. In this the Parke-Taylor 
expressions for MHV amplitudes~\cite{ParkeTaylor},
\vspace{-0.3cm}
$$
\hspace{-0cm}\eqalign{
\Atree_{[1^+,\ldots,j^-,\ldots,k^-,
                \ldots,n^+]}\,
\!= i { {\spa{j}.{k}}^4 \over \spa1.2\spa2.3\cdots\spa{n}.1 }\,,
\cr}
\hspace{-.4cm}\equn\label{ParkeTaylor}
\vspace{-0.2cm}
$$ 
are promoted to vertices for a diagrammatic expansion~\cite{CSW}. 
The twistor variables for the 
intermediate momentum between vertices are calculated through the relation 
$\lambda_a(q) =  q_{a\dot a} \eta^{\dot a}$
with $\eta$ being a reference spinor. 
Since the Parke-Taylor expression only involves holomorphic 
variables $\lambda_a$ and not anti-holomorphic variables 
$\bar\lambda_{\dot a}$ this is sufficient to define the expansion.
Gravity MHV amplitudes~\cite{BerGiKu} involve both $\lambda_a$ and
$\bar\lambda_{\dot a}$ and it has proven difficult to find the correct
continuation of $\bar\lambda_{\dot a}$~\cite{GBgravity}. Despite this
in ref.~\cite{BeBbDu}, it was demonstrated that gravity amplitudes
satisfy the same type of localisation properties in twistor space as
Yang-Mill amplitudes. More recently it was shown in ref.~\cite{gravCSW} 
that a MHV construction was possible provided that one carries out 
special shifts~\cite{Kasper} in the anti-holomorphic 
variables $\bar\lambda_{\dot a}$. 

Gravity amplitudes are also amenable to a BCFW-like shift as 
demonstrated in refs.~\cite{BCFWgravity}.

Thus, although more complicated, the new techniques may be 
applied to amplitudes of theories with gravity. 

\section{One-Loop in $\NeqEight$ Supergravity}
At one-loop level, string theory would suggest that the 
KLT relations extends {\it within} the loop momentum integrals. 
After integration, however, such relations 
would not be expected to persist in the amplitude. 
To illustrate this we examine the one-loop
amplitudes in maximal supergravity/Yang-Mills.  
In evaluating loop amplitudes one performs integrals 
over the loop momenta, $\ell^\mu$, with polynomial 
numerator $P(\ell^\mu)$. In a Yang-Mills theory, the loop 
momentum polynomial will generically be of degree $\leq n$ for a 
$n$-point loop. $\NeqFour$ one-loop amplitudes exhibits 
considerable simplification and the loop momentum 
integral will be of degree $n-4$~\cite{StringBased,BDDKa}. 
Consequently, the amplitudes can be expressed as a sum of scalar box 
integrals with rational coefficients, as follows from a Passarino-Veltman 
reduction~\cite{PassVelt},
\vspace{-0.1cm}
$$
\Aloop= \sum_{a}  c_a I^4_a\,.\equn\label{OnlyBoxesEQ}
\vspace{-0.2cm}
$$
Considerable progress has recently been made in determining 
such coefficients, $ c_a$, using a variety of methods 
based on unitarity~\cite{BDDKa,BDDKb,NeqFour}. 

For maximal $\NeqEight$ supergravity~\cite{ExtendedSugra}
the equivalent power counting arguments~\cite{GravityStringBased} 
give a loop momentum polynomial of degree
\vspace{-0.1cm}
$$
2(n-4)\,,
\equn
\vspace{-0.2cm}
$$ 
which is consistent with eq.~(\ref{StringRelation}).   
Reduction for $n>4$ leads to a sum of tensor box integrals 
with integrands of degree $n-4$ which would then reduce
to scalar boxes {\it and} triangle, bubble and rational 
functions,
\vspace{-0.1cm}
$$
\Mloop= \sum_a c_a I_4^a +\sum_a d_a I^a_3 +\sum_a e_a I^a_2 +R
\equn\label{gravityloopEQ}
\vspace{-0.2cm}
$$
where the $I_3$ are present for $n\geq 5$, $I_2$ for $n \geq 6$ and the
rational terms for $n\geq 7$. 

The first  calculation of an one-loop $\NeqFour$ amplitude was 
of the four point~\cite{GSB}, 
\vspace{-0.1cm}
$$
\Aloop_{[1,2,3,4]}
=st \times \Atree_{[1,2,3,4]} \times I_{4\,(s,t)}\,.
\equn
\vspace{-0.2cm}
$$ 
Here $I_{4, (s,t)}$ denotes the scalar box integral with attached 
legs in the order $1234$ and $s$, $t$ and $u$ are the usual 
Mandelstam variables. The $\NeqEight$ amplitude was also given,
\vspace{-0.1cm}
$$
\eqalign{
\Mloop_{[1,2,3,4]}&=stu \Mtree_{[1,2,3,4]}
\Bigl[ I_{4\,(s,t)}\!+\!I_{4\,(s,u)}\!+\!I_{4\,(t,u)}
\Bigr]\,,}\equn
\vspace{-0.2cm}
$$
so that, like the $\NeqFour$ Yang-Mills amplitude, the $\NeqEight$
amplitude can be expressed in terms of scalar box-functions. 
For $n=4$ this similarity between $\NeqFour$ and $\NeqEight$ is consistent
with the previous power counting arguments. 

 \section{The {\it no-triangle} hypothesis} 
Despite the power counting argument, there is evidence~\cite{BeBbDu} 
that one-loop amplitudes of $\NeqEight$ can be expressed simply 
as a sum over scalar box integrals analogous to the $\NeqFour$
amplitudes~(\ref{OnlyBoxesEQ}).  
We label this as the ``no-triangle hypothesis.''  
We emphasis that this is a hypothesis: we will present evidence in 
its favour but not a proof. 

Firstly, in the few definite computations at one-loop level, triangle
or bubble functions do not appear.  In ref.~\cite{Bern:1998sv} the
five and six point ``Maximally Helicity Violating'' amplitudes were
computed and contrary to expectations, consisted entirely of scalar
box-functions.

Secondly, factorisation properties of the physical amplitudes 
do not seem to demand the presence of these functions. Since the
four and five point amplitudes are triangle-free then in any soft, 
collinear, or multi-particle pole limit of a higher point
function the triangles would by necessity drop out or be absent 
in the first place. For further support we refer to ref.~\cite{Bern:1998sv}, 
where an ansatz for the $n$-point MHV amplitude was constructed, entirely of 
box functions consistent in all soft limits.
The simplification is peculiar to $\NeqEight$ and does not 
apply for ${\cal N} < 8$ supergravities~\cite{GravityStringBased,Dunbar:1999nj}.

Thirdly, one can calculate the box-coefficients for the amplitude
using unitarity and examine whether the amplitude has the correct soft
behaviour. At one-loop, the expected soft divergence in a $n$
graviton amplitude is~\cite{DunNorB},
\vspace{-0.1cm}
$$\small
\eqalign{
M^{\rm one-loop}_{
[1,2,\ldots, n]}
=
{i c_\Gamma \kappa^2}
\bigg[
{\sum_{i<j}  s_{ij} \ln[-s_{ij} ]
\over 2\epsilon}
\bigg]
\!\!\times\! M^{\rm tree}_{[1,2,\ldots, n]}.
\cr}
\equn\label{IRamplitudeEQ}
\vspace{-0.15cm}$$ 
In the expansion of the one-loop amplitude~(\ref{gravityloopEQ})
both box and triangle integral functions contain divergences of the form
$\ln(-P^2)/\eps$ and the bubble integrals contains $1/\eps$
divergences.  Thus if the boxes contain all the correct IR divergences
we can conclude that the remaining parts cannot contribute any IR
divergences. The triangle integrals can be organised according to the
number of ``massive'' legs. The one and two mass triangles are not all
independent but choosing an appropriate subset as a basis we can
immediately deduce that these must be absent from the amplitude when the
boxes correctly contain the IR singularities. (A caveat is that the
three-mass triangle contains no IR singularity and so cannot be
excluded by an IR argument.)  In ref.~\cite{BBDuIt} the box
coefficients were explicitly computed for the six-point NMHV
amplitudes. The result consisted of a sum of one-and two-adjacent-mass 
boxes:
\vspace{-0.1cm}
$$
\eqalign{
\Mloop_{[a,b,c,d,e,f]}&=
c_{\Gamma} 
\Big( 
\hspace{-0.6cm}\sum_{(abcdef)\in P_6''} \hspace{-0.5cm} \hat c^{\,(abc)def} I_4^{(abc)def}
\cr
&+\hspace{-0.4cm}
\sum_{(abcdef)\in P_6'}  \hspace{-0.5cm}\hat c^{\,a(bc)(de)f} I_4^{a(bc)(de)f}
\Big)
\,.
\cr}\equn
\vspace{-0.15cm}
$$ 
The six-point amplitude thus calculated gives exactly
the entire expected IR structure of the one-loop amplitude confirming the
absence of triangles (with at least one massless leg). 

The ``no-triangle'' hypothesis applies to one-loop
amplitudes. However, by factorisation it conceivably extends
beyond one-loop. 
We wish to comment that the hypothesis, 
if true, implies a significantly softer UV behaviour of gravity 
than expected from power counting.  

\section{Multi-loop amplitudes}

\def\P{P}
\def\NP{NP}

The two-loop four-point amplitudes are~\cite{Bern:1997it},
\vspace{-0.1cm}
$$
\eqalign{ 
&{ A}_4^{\twoloop} = 
g^6
 st  
 \,A_4^{\rm tree}
 \Bigl[s \, \I_{4\,(t, s)}^{\twoloop,\P}
\!+\!s\,  \I_{4\,(t, s)}^{\twoloop,\P} 
\Bigr]
\cr}
\hspace{-1cm}\equn
\vspace{-0.15cm}
$$
and~\cite{BDDPR},
\vspace{-0.1cm}
$$
\eqalign{ 
{\cal M}_4^{\twoloop}  &= 
 \Bigl({\kappa \over 2} \Bigr)^6 \, stu \, 
 M_4^{\rm tree}\cr &\hspace{-.3cm}\!\times\!
 \Bigl[s^2\, \I_{4\,(s, t)}^{\twoloop,\P}  
\!\!+ s^2\, \I_{4\,(s, t)}^{\twoloop,\NP}
\!\!+\! {\rm perms} \Bigr] \,, \cr}
\hspace{-1cm}\equn
\vspace{-0.15cm}
$$ where $P$ and $NP$ indicate the planar and non-planar two-loop
scalar box functions.  The amplitudes are both UV finite for $D\leq 6$. 
A thorough discussion of the power counting of
amplitudes with $L>2$ for $D\geq 4$ was presented in
ref.~\cite{BDDPR}.  By examining the cuts of higher point
functions the leading UV behaviour of  $\NeqFour$ SYM is expected to be,
\vspace{-0.1cm}
$$
\int (d^D p)^L  {(p^2)^{(L-2)}  \over (p^2)^{3L +1}}  \,,
\equn
\vspace{-0.15cm}
$$ implying amplitudes to be finite when
\vspace{-0.1cm}
$$
\eqalign{
   D & < {6 \over L }+ 4 \;. 
\cr}
\vspace{-0.1cm}
$$  
The expectation for three-loop
Yang-Mills has recently been confirmed in~\cite{ThreeLoopYM}. For $D=4$
this is restating the known finiteness of  $\NeqFour$ SYM~\cite{Mandelstam}.

For supergravity, the expectation is that the integrands will be the
square of Yang-Mills integrands and thus the divergence will go as,
\vspace{-0.1cm}
$$
\int (d^D p)^L  {(p^2)^{2(L-2)}  \over (p^2)^{3L +1}}  \,,
\equn
\vspace{-0.1cm}
$$
implying finiteness when
\vspace{-0.1cm}
$$
\eqalign{
  D &< { 10 \over  L}+2\,.
\cr}
\vspace{-0.1cm}
$$ This suggest that $\NeqEight$ supergravity is infinite at five loops in
$D=4$. Explicitly, this behaviour was checked for parts of the amplitude  which are ``two-particle cut-constructible''.  Although we do not have a
candidate mechanism for the no-triangle hypothesis, it does suggest a
cancellation {\it between} diagrams. Such cancellations, if they persist
beyond one-loop would suggest the above power counting is too
conservative. 
 
\section{Conclusions}
The recent progress in computing gauge theory amplitudes can be
extended, in many cases, to theories incorporating gravity. This has
improved our understanding of the perturbative expansion of, in
particular, maximal $\NeqEight$ supergravity.

The current status on the perturbative expansion is that $\NeqEight$ 
is two loop finite, but expected to diverge from power counting 
arguments at five loops~\cite{Howe:2002ui,BDDPR}. However it is surprising 
that in concrete calculations~\cite{BDDPR,Bern:1998sv,BeBbDu,BBDuIt} 
the large momentum structure in $\NeqEight$ supergravity appears 
to be much simpler than power counting suggests. These simplifications 
are completely unexpected from the currently 
known symmetries of $\NeqEight$ supergravity.
One might suspect, this implies the existence of further symmetries 
and additional constraints on the scattering amplitudes. It seems promising,
although challenging, to utilise the simplification as well as new techniques 
to determine the ultra-violet behaviour of higher loop
scattering amplitudes in $\NeqEight$ supergravity.

\end{document}